\documentclass[sigconf,screen]{acmart}
\setcopyright{none}

\AtBeginDocument{%
  \providecommand\BibTeX{{%
    \normalfont B\kern-0.5em{\scshape i\kern-0.25em b}\kern-0.8em\TeX}}}

\setcopyright{acmcopyright}
\copyrightyear{2022}
\acmYear{2022}
\acmDOI{}

\acmConference[FAccTRec '22]{}{}{Seattle, WN}
\begin{document}

\title{Inclusive Ethical Design for Recommender Systems}

\author{Susan Leavy}
\email{susan.leavy@ucd.ie}
\orcid{0000-0002-3679-2279}
\affiliation{%
  \institution{Insight SFI Research Centre for Data Analytics\\
  School of Information and Communication Studies\\
  University College Dublin\\}
  \state{Dublin}
  \country{Ireland}
}

\renewcommand{\shortauthors}{Leavy S.}

\begin{abstract}

Recommender systems are becoming increasingly central as mediators of information with the potential to profoundly influence societal opinion. While approaches are being developed to ensure these systems are designed in a responsible way, adolescents in particular, represent a potentially vulnerable user group requiring explicit consideration. This is especially important given the nature of their access and use of recommender systems but also their role as providers of content. This paper proposes core principles for the ethical design of recommender systems and evaluates whether current approaches to ensuring adherence to these principles are sufficiently inclusive of the particular needs and potential vulnerabilities of adolescent users. 

\end{abstract}


\maketitle

\section{Introduction}

Recommendation algorithms are possibly the most pervasive and influential AI systems in use today. They augment user's behaviour, beliefs and preferences \cite{carroll2022estimating,ashton2022recognising,evans2021user} thereby having the capacity to induce widespread change in society. The influence of recommendation algorithms are often seen as benign, pertaining to entertainment choices and simply mirroring changes in user preferences. However, there is now, mounting evidence of the potential serious harms of recommender systems~\cite{allcott2020welfare,shakya2017association}.

Although levels of proliferation of a few popular recommendation systems presents potential for widespread social change, public perception of the risk of Artificial Intelligence (AI) systems are not focused on such change. Rather, perceptions of AI systems among tend to centre on risks to privacy, job loss or threats to human autonomy \cite{cave2019scary,kieslich2021threats}. Fears of threats to human autonomy are characterised, particularly in the media, as emanating from the potential emergence of sentient AI beings \cite{hermann2021artificial,nader2022public}. However, unlike the common portrayal of recommender systems as being designed to present relevant content to users, as demonstrated by Carroll et al.\cite{carroll2022estimating} they an also learn to change humans with regards to their preferences and subsequent behaviour and it follows therefore that the undermining of human autonomy may already be taking place. 

That recommender systems may influence, not only how people behave, bunt their interests, is a particularly urgent issue in the context of vulnerable users. If recommendation systems for instance, have the ability to change what children are interested in, in ways that may not be foreseen by designers of these systems, in order that they remain engaged consumers of content then this presents profound implications for large-scale AI-driven transformation of societal opinion. While some work is ongoing to address the particular needs of children with respect to AI systems, this work primarily focuses on young children. Adolescents however, are being given smartphones at increasingly early ages and often with little adult insight or oversight \cite{vaterlaus2021smartphone, anderson2018teens,akter2022parental}. They are therefore being increasingly treated as adults in a context where recommendation systems, by virtue of their aim to increase user engagement, have the capacity to fundamentally alter their opinions, preferences and behaviour in ways neither they, their guardians or system designers can foresee.

\section{Inclusive Ethical Design}

Given that adolescents increasingly, have the same level of access as adults online, it follows that proposed ethical principles for the design and development of recommender systems must be inclusive of them. Frameworks for ethical design and development of AI systems must include adolescents as potential users of recommender systems from an intersectional perspective considering aspects of identify such as race, gender and socio-economic context.  Proposed EU regulation does highlight children as particularly vulnerable users warranting special protection and they propose the \emph{`prohibition of practices that have a significant potential to manipulate persons through subliminal techniques beyond their consciousness or exploit vulnerabilities of specific vulnerable groups such as children or persons with disabilities in order to materially distort their behaviour in a manner that is likely to cause them or another person psychological or physical harm}'\cite{act2021proposal}. Given this, it is incumbent on designers, auditors and regulators of AI systems to explicitly consider such vulnerable users. 

In exploring whether existing ethical frameworks for AI systems encompasses potential adolescent users, this paper builds on work by Leavy et al.~\cite{leavy2021ethical} that proposes a frame for the evaluation of AI informed by feminist epistemology and critical theories of race. This paper proposes ethical principles for recommender systems and evaluates whether current approaches to implementing ethical design are sufficiently inclusive of adolescents as users.


\subsection{Including Diverse Perspectives}
 
The perspectives of those involved in the design and development of AI systems are embedded in their design and can subsequently influence the social construction of concepts in society~\cite{costanza2018design, ahsen2019algorithmic}.  It is important therefore as part of ethically evaluating recommender systems, that the perspectives of those involved in their design and developed are critically evaluated.  According to the AI Now Institute~\shortcite{west2019discriminating}, the dominant group whose values are most likely to be captured within AI systems are straight, white, cisgender, college-educated men from a few of the largest technology companies. 

In an attempt to address this issue, there are several initiatives to promote diversity among the perspectives represented in the development of AI. However, given legal status as children among other factors, there are no initiatives to increase the representation of adolescents in the tech industry. Current drives to increase diversity in the workforce therefore can not address the lack of representation of the perspectives of adolescents in the design of recommender systems. Given the absence of adult oversight of their interaction with recommender systems and the effect we know they could have on them, the absence of their involvement leaves them vulnerable to the motivations of those who may wish to capitalise from their ever increasing engagement to the detriment of their well being.

\subsection{Recognising Reflexivity}
Choices in how knowledge is represented can generate and augment concepts in society. Recommender systems, by virtue of their function as gatekeepers and curators of knowledge, particularly in representations of the world through media, have a particularly influential role in how fundamental societal concepts are framed. this is for instance, particularly pertinent in the case of gender and race. Judith Butler~\shortcite{butler1988performative} for instance, spoke of how humans (re)produce gender through language and how it is used. The mediation of this process through AI algorithms however, adds an entirely new dimension to the poststructuralist view of the reflexive process of the generation of core societal concepts. 

Adopting a feminist viewpoint would necessitate an activist stance to ensure that concepts are being re-produced and represented in alignment with ethical values. One example of how this may be implemented is by the curation of training data for AI in an ethical way, mitigating the  risk of  patterns of social injustice being learned from historical data. This is what what Viginia Dignum~\shortcite{dignum2019responsible} described  as curating data in a way that captures `the world as it should be'. While there has been work on adolescents and how social media use affects them, in particular in the context of teenage girls, the role of recommender systems in (re)producing societal concepts and curating their presentation to adolescents in a highly personalised way, with the goal of ensuring online engagement, has however, not been fully addressed within current research. 


\subsection{Understanding Theory in Systems}

Philosophical viewpoints can be deeply embedded in how knowledge and information is represented in data and in the media in particular~\cite{gillborn2018quantcrit,borgman2015big,d2020data}. The value-laden nature of human knowledge and how this filters from data that machine learning algorithms are trained upon, through to AI systems has been widely critiqued. Leavy et al.\cite{leavy2021ethical} propose an examination from a feminist and anti-racist standpoint, the theoretical view of race and gender embedded in AI systems. Adolescents themselves are increasingly authors of media content that is in turn mediated by recommender systems. The dynamics of this mediation in the context of adolescents has the potential to profoundly impact the formation of their own theoretical standpoints. This is evident in how user preferences shift as a  result of interaction with recommender systems~\cite{carroll2022estimating}. The effect may be compounded further given the self-generated nature of much of the content that is mediated for adolescents.

\subsection{Include Marginalised Groups}

There have been calls to include within the design process of AI systems,  perspectives and forms of knowledge from groups that are marginalised through development of participatory approaches to AI design~\shortcite{costanza2018design,simonsen2012routledge}. These approaches however generally assume participants are adults. While the scale of adolescent use of social media platforms  presents an opportunity to engage with the \emph{``ideas, and feelings of hundreds of millions of people''}~\cite{manovich2011trending}. Given the central role of adolescents in creating content however,  their lack of voice or input into the design of the systems that curate it, select who will view it or how it is profited from, adolescents are arguably a highly vulnerable and potentially exploited group. An ethical design of recommender systems therefore necessitates the development of standards whereby perspectives of adolescents can be explicitly included in the design process.

\section{Conclusion}

While there is some research evaluating the impact of technology on adolescent well-being, there is little specifically highlighting how societal concepts are mediated by recommender systems and the effect this may have on them. Given their status as vulnerable groups, this paper proposed ethical principles for the design of recommender systems and highlighted how adolescent groups in particular need to be incorporated explicitly. Currently, while adolescents are increasingly treated as adults in online spaces there is no currently pathway for their perspectives to be incorporated into the design of AI systems. Is is crucial therefore that approaches to ensure ethical recommender systems are broadened to be more inclusive of adolescents.

\section*{Acknowledgements}
This publication has emanated from research conducted with the financial support of Science Foundation Ireland under Grant number 12/RC/2289\_P2. 
For the purpose of Open Access, the author has applied a CC BY public copyright licence to any Author Accepted Manuscript version arising from this submission.

\bibliographystyle{ACM-Reference-Format}

\end{document}